%% file: veriprop23.tex
\documentclass[runningheads]{llncs}
\usepackage[T1]{fontenc}
\usepackage{graphicx}
\usepackage[margin=.9in]{geometry}
\usepackage{xcolor}
\usepackage{amsmath}
\usepackage{listings}
\usepackage{inconsolata}
\usepackage{wrapfig}
\usepackage{hyperref}
\usepackage{cleveref}
\usepackage{microtype}

\begin{document}
\title{Verifying Performance Properties of Probabilistic Inference\vspace{-.5em}}

\author{Eric Atkinson\inst{1} \and
Ellie Y. Cheng\inst{1} \and
Guillaume Baudart\inst{2} \and
Louis Mandel\inst{3} \and
Michael Carbin\inst{1}}
\authorrunning{E. Atkinson et al.}
\institute{Massachussetts Institute of Technology, USA \and
\'{E}cole normale sup\'{e}rieure, PSL University, CNRS, Inria, France \and
IBM Research, USA}
\maketitle              % typeset the header of the contribution

\newcommand{\approxs}[0]{\textsc{approx}}
\newcommand{\exacts}[0]{\textsc{exact}}
\newcommand{\approxa}[0]{\lstinline{approx}}
\newcommand{\exacta}[0]{\lstinline{exact}}

%\begin{abstract}\todo{XXX}
%%\keywords{First keyword  \and Second keyword \and Another keyword.}
%\end{abstract}
%
%
\input{intro}

\input{ssi}

\input{ci}

\input{static_ds}

\input{conclusion}

%
%
%\subsubsection{Acknowledgements} \todo{XXX} 
%\vspace{-.5em}
\bibliographystyle{splncs04}
\begingroup
\renewcommand{\section}[2]{}%
{\large\bibliography{veriprop23}}
\endgroup

\end{document}

%% file: intro.tex
\vspace{-2em}
\subsubsection{Introduction.}

Probabilistic inference is an NP-hard problem~\cite{cooper1990computational}, meaning probabilistic inference systems can have poor performance in general.
Nevertheless, efficient inference techniques exist for many special classes -- including linear-Gaussian models~\cite{kalman}, Bayesian networks that are polytrees~\cite{polytree}, and Monte Carlo sampling techniques that work well in widespread practical applications.
\emph{Probabilistic Programming Languages (PPLs)} provide a general interface for developers to write a program to specify nearly any probabilistic inference problem.
This raises the question: how can developers using PPLs have confidence that their programs will achieve good performance (e.g.\ by falling into one of the special classes)?

In this abstract, we discuss the opportunity to formally verify that inference systems for probabilistic programming guarantee good performance.
In particular, we focus on \emph{hybrid inference} systems that combine exact and approximate inference to try to exploit the advantages of each. Their performance depends critically on a) the division between exact and approximate inference, and b) the computational resources consumed by exact inference.
We describe several projects in this direction:\vspace{-.5em}
\begin{itemize}
\item \emph{Semi-symbolic Inference (SSI)}~\cite{semisymb_oopsla22} and \emph{Delayed Sampling (DS)}~\cite{murray_ds} are types of hybrid inference systems, with SSI providing limited guarantees by construction on the exact/approximate division.
\item \emph{Verifying the Exact/Approximate Division} is ongoing work to extend SSI's guarantees to a more complex class of programs, requiring a program analysis to ensure the guarantees.
\item \emph{Verifying Memory Consumption} is prior work~\cite{probzelus_analysis} on verifying that DS inference systems execute in bounded memory.
\end{itemize}
\vspace{-.5em}Together, these projects show that verification can deliver the performance guarantees that PPLs need.

%% file: ssi.tex
\newcommand{\sample}{{\ensuremath{\leftarrow}}}
\begin{wrapfigure}[15]{r}{0.4\textwidth}
\vspace{-2.5em}
\begin{lstlisting}
function outlier(yobs) {
  x *\sample* gaussian (0., 100.);*\label{code:latent1}*
  for i in 1 .. N {
    x *\sample* gaussian(x, 1.);*\label{code:latent2}*
    o *\sample* bernoulli(.1);*\label{code:outlier}*
    if (o) { y *\sample* gaussian(0., 100.); }*\label{code:obs_outlier}*
    else   { y *\sample* gaussian(x, 1.); }*\label{code:obs_normal}*
    observe(y, yobs[i]);
  }; x*\label{code:return}*
}
\end{lstlisting}
\caption{The \emph{outlier} example. Note that we use \sample\ to sample from a distribution and assume the input \lstinline*yobs* has length at least \lstinline*N*.}
\label{fig:outlier}
\end{wrapfigure}

\vspace{-1em}
\subsubsection{Semi-Symbolic Inference and Delayed Sampling.}

SSI~\cite{semisymb_oopsla22} and its predecessor DS~\cite{murray_ds} are instances of hybrid inference where the runtime performs exact inference on known efficient special classes with symbolic distributions, and falls back on general approximate sampling when necessary.

Consider the \emph{outlier} example (pseudocode in Fig~\ref{fig:outlier}, adapted from~\cite[Section~2]{Minka01}).
This models a latent random variable \lstinline*x* that evolves according to a Gaussian random walk (Lines \ref{code:latent1} and \ref{code:latent2}). Observations can be outliers with probability 10\% (Line~\ref{code:outlier}). At each step with an outlier, the observation \lstinline*y* is modeled with a Gaussian unrelated to \lstinline*x* (Line~\ref{code:obs_outlier}). Without an outlier, \lstinline*y* is modeled with a Gaussian distribution with mean \lstinline*x* as in a standard Kalman filter~\cite{kalman} (Line~\ref{code:obs_normal}). The code returns \lstinline*x* (Line~\ref{code:return}), indicating inference of \lstinline*x* from the last iteration, conditioned on all prior observations of \lstinline*y* as values in the array \lstinline*yobs*.

For this program, the goal for SSI and DS during inference is to approximately sample values for \lstinline{o} at each step.
Then, given concrete values for \lstinline{o}, the symbolic distributions for the remaining variables form a linear-Gaussian model that is amenable to exact inference.
While both SSI and DS can implement this system on the \emph{outlier} example, one advantage of SSI is that it provides a guarantee: given \emph{any} program where each variable is either a) approximate, or b) linear-Gaussian, SSI will never approximate any of the linear-Gaussian variables (see \cite{semisymb_oopsla22} for details, as well as examples where DS fails to provide this guarantee).
This ensures that, under limited circumstances, SSI achieves the optimal division between exact and approximate inference.

%% file: ci.tex
\vspace{-1em}
\subsubsection{Verifying the Exact/Approximate Division.}

Despite its guarantees, SSI's division between exact and approximate inference within a program remains enigmatic.
For example, in the \emph{outlier} program, depending on the internal structure of the symbolic state, SSI may choose to approximate \lstinline{o} and compute \lstinline{x} and \lstinline{y} using exact inference, or alternatively approximate all variables \lstinline{o}, \lstinline{x}, and \lstinline{y}.
In general, determining which random variables SSI will choose to approximate can require in-depth knowledge of the algorithm, and the wrong choice can significantly degrade performance~\cite{rppl,semisymb_oopsla22}.

We propose the use of annotations to control the division of exact and approximate inference at the granularity of individual random variables. Developers force the runtime to approximate a random variable using the \approxa{} annotation (this is similar to the \lstinline*value* construct from prior work~\cite{murray_ds,rppl,semisymb_oopsla22}). In the \emph{outlier} example, annotating \lstinline{o} with \approxa{} would cause SSI to always approximate \lstinline{o}, guaranteeing that SSI would perform exact inference on \lstinline{x} and \lstinline{y}. 
A developer could further apply the \exacta{} annotation to \lstinline{x}, which functions as an assertion that the runtime will compute \lstinline{x} exactly.
Conversely, should the user fail to annotate \lstinline{o} with \approxa{}, then an \exacta{} annotation on \lstinline{x} will raise an error.

We further propose to formally verify \exacta{} annotations at compile-time, and automate this verification with an abstract-interpretation-based static analysis.
Such an analysis would be provably sound -- i.e.\ never consider a variable exact that SSI approximates in any execution -- and we hypothesize it will empirically be precise enough for common examples.
Thus, the analysis provides a certificate that a variable will be exact, which serves as an enhanced version of SSI's performance guarantee.

%% file: static_ds.tex
\vspace{-1em}
\subsubsection{Verifying Memory Consumption.}
For both SSI and DS, the exact inference component of the system maintains a symbolic distribution representation at runtime.
This leads to the question: how large can this representation grow?
Developers would like their inference runtimes to maintain \emph{bounded memory}, meaning the size of the runtime state is a constant multiple of the number of variables in the program (see~\cite{probzelus_analysis} for a detailed definition).
For example, in the \emph{outlier} example of Fig.~\ref{fig:outlier}, although each iteration instantiates three \emph{new} random variables in the symbolic state (assigning them to the program variables \lstinline{x}, \lstinline{o}, and \lstinline{y}) only a \emph{total} of three random variables are ever need at runtime (the ones pointed to by the program variables). In this case, a bounded-memory runtime would use a constant amount of memory regardless of the value of \lstinline{N}.

Whether or not DS runs in bounded memory depends on the probabilistic program in question~\cite{rppl,probzelus_analysis}.
The work in \cite{probzelus_analysis} identified two program properties -- the $m$-consumed property and the unseparated paths property -- which together form a necessary and sufficient condition for DS to be bounded-memory.
These are dataflow properties of the program that are unique to the bounded-memory inference problem.
The work in \cite{probzelus_analysis} also presents how to automatically verify these properties with a static analysis that is provably sound and empirically -- across a range of benchmarks -- precise enough to verify bounded-memory execution of numerous programs.
Overall, this provides a provably sound way to ensure DS inference systems achieve good memory performance (we are not yet aware of any work extending this to SSI).

%% file: conclusion.tex
\vspace{-1em}
\subsubsection{Conclusion.}

We have discussed techniques to formally verify that hybrid inference systems will achieve good performance on a program.
We discussed SSI, its guarantee on the exact/approximate division, and ongoing work to extend this guarantee.
We also discussed how to formally verify DS runs in bounded memory.

We close by proposing several directions for future work.
A natural direction is to extend the DS bounded-memory guarantees to SSI.
We hypothesize that the same DS properties are sufficient but not necessary for bounded-memory SSI, suggesting that this endeavor will require more work.
A related direction is to determine bounds on SSI's computational runtime to complement the memory bounds.
We suspect computation bounds to be related to tree-width as is the case for Bayesian networks.
Finally, one of the major sources of performance unpredictability in hybrid inference is the amount of approximation error.
While reasoning about the error from approximate inference is a challenging problem, we predict that recent work on efficiency analysis of rejection sampling~\cite{bn_rejection_efficiency} can be adapted to formally verify the approximation error of hybrid inference systems.
Together, tools for verifying the exact/approximate division, memory consumption, computational runtime, and approximation error would provide important guarantees for most if not all sources of performance issues in hybrid inference systems for probabilistic programming.